\documentclass[sigconf,nonacm]{acmart}

\usepackage{graphicx}

\usepackage{amsmath}
\usepackage{algorithm,algorithmic,color, multirow}
\usepackage{bm}
\usepackage{booktabs}
\usepackage{subcaption}
\usepackage{caption, amsmath}
\usepackage{graphicx}
\usepackage{soul}

\begin{document}

\title{Distributed Federated Learning for Vehicular Network Security: Anomaly Detection Benefits and Multi-Domain Attack Threats}

\author{Utku Demir}
\affiliation{
  \institution{Nexcepta Inc.}
  \city{Gaithersburg}
  \state{MD}
  \country{USA}
  \postcode{20855}}

\author{Yalin E. Sagduyu}
\affiliation{
  \institution{Nexcepta Inc.}
  \city{Gaithersburg}
  \state{MD}
  \country{USA}
  \postcode{22203}}

\author{Tugba Erpek}
\affiliation{
 \institution{Nexcepta Inc.}
  \city{Gaithersburg}
  \state{MD}
  \country{USA}}
  \postcode{22203}

\author{Hossein Jafari}
\affiliation{
 \institution{Nexcepta Inc.}
  \city{Gaithersburg}
  \state{MD}
  \country{USA}
  \postcode{22203}}

\author{Sastry Kompella}
\affiliation{
 \institution{Nexcepta Inc.}
  \city{Gaithersburg}
  \state{MD}
  \country{USA}
  \postcode{22203}}

\author{Mengran Xue}
\affiliation{
 \institution{RTX BBN Technologies}
  \city{Cambridge}
  \state{MA}
  \country{USA}
  \postcode{22203}}


\begin{abstract}
In connected and autonomous vehicles, machine learning for safety message classification has become critical for detecting malicious or anomalous behavior. However, conventional approaches that rely on centralized data collection or purely local training face  limitations due to the large scale, high mobility, and heterogeneous data distributions inherent in inter-vehicle networks. To overcome these challenges, this paper explores Distributed Federated Learning (DFL), whereby vehicles collaboratively train deep learning models by  exchanging model updates among one-hop neighbors and propagating models over multiple hops. Using the Vehicular Reference Misbehavior (VeReMi) Extension Dataset, we show that DFL can significantly improve classification accuracy across all vehicles compared to learning strictly with local data. Notably, vehicles with low individual accuracy see substantial accuracy gains through DFL, illustrating the benefit of knowledge sharing across the network. We further show that local training data size and time-varying network connectivity correlate strongly with the model’s overall accuracy. We investigate DFL's resilience and vulnerabilities under attacks in multiple domains, namely wireless jamming and training data poisoning attacks. Our results reveal important insights into the vulnerabilities of DFL when confronted with multi-domain attacks, underlining the need for more robust strategies to secure DFL in vehicular networks. 
\end{abstract}




\maketitle
\pagestyle{plain}
\section{Introduction}\label{Intro}
The rapid evolution of connected and autonomous vehicles (CAVs) is reshaping modern transportation systems. Recent years have witnessed significant advancements in inter-vehicle communication, spurred by the emergence of connected and autonomous vehicles and the increasing demand for real-time data exchange. These developments, facilitated by protocols such as Dedicated Short-Range Communications (DSRC) and Cellular Vehicle-to-Everything (C-V2X), have paved the way for various applications in cooperative safety, traffic optimization, and entertainment services \cite{9645261}. As vehicles increasingly rely on real-time data sharing and cooperative decision-making, ensuring the reliability and security of safety message exchange has become paramount. Anomaly detection, namely identifying malicious messages received by individual vehicles, is critical to preventing cascading failures and ensuring road safety. However, traditional anomaly detection approaches, which rely on centralized data aggregation or isolated local training, struggle to address the complexities of dynamic vehicular environments and remain insufficient in the face of highly mobile network topologies, geographically dispersed data, and stringent latency constraints.

Federated learning (FL) has emerged in vehicular networks as a means of leveraging the abundant distributed data generated by onboard sensors (e.g., LiDAR, radar, cameras) and network data (e.g., beacon messages). By training a shared model over multiple decentralized nodes, FL reduces the need for raw data transfer to a central server and offers privacy benefits, as only local model updates are communicated. FL finds various applications in vehicular networks. For security purposes, FL was applied for misbehavior detection of safety messages \cite{huang2024semi}, anomaly detection in the Internet of Vehicles \cite{tham2023federated}, vehicle trajectory prediction against cyber attacks \cite{wang2023federated} and V2X misbehavior detection in 5G edge networks \cite{yakan2023federated}.

Despite its benefits, most existing FL implementations still rely on a central server that aggregates and redistributes model parameters. Although effective in stable or low-mobility scenarios, this architecture encounters practical challenges in vehicular networks, where connectivity is transient and fixed infrastructure may be unavailable. 

\begin{figure}[ht!]
	\centering
	\includegraphics[width=\columnwidth]{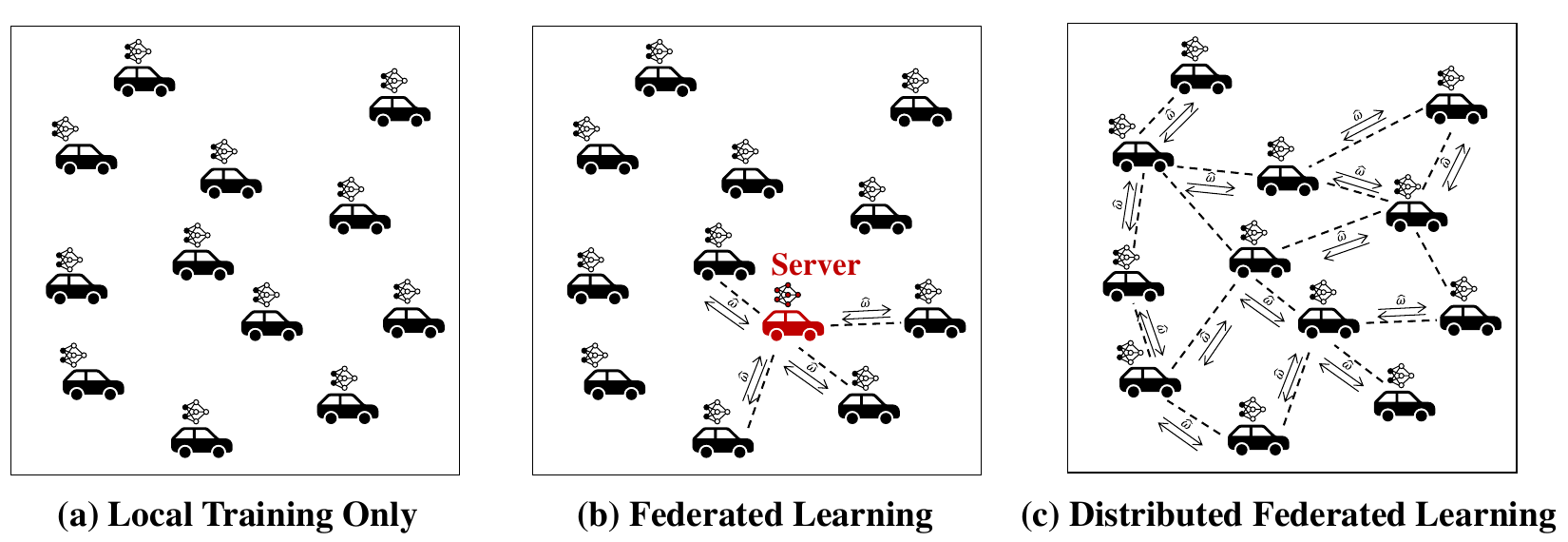}
	\caption{Different modes of learning: (a) local training only, (b) federated learning, (c) distributed federated learning.}
	\label{fig:DFL}
\end{figure}

These limitations have prompted us to explore Distributed Federated Learning (DFL), which eliminates the reliance on a single central aggregator in favor of multiple local aggregations among neighboring vehicles \cite{shi2022federated}. DFL leverages the computational resources of individual vehicles while preserving data privacy, and enables vehicles to propagate updates through a multi-hop network, overcoming limitations associated with centralized systems by allowing faster adaptation and reducing single-point-of-failure risks. In DFL, vehicles engage in direct, one-hop model exchanges that propagate aggregated knowledge over multiple hops enhancing scalability and improving resilience against connectivity disruptions. Different modes of learning are illustrated in Figure~\ref{fig:DFL}. 

DFL in vehicular networks faces several challenges. The dynamic topology and variable connectivity of inter-vehicle networks make consistent and timely model aggregation a complex task. Synchronization issues can emerge when nodes differ in connectivity or local data volume, potentially hampering the convergence and stability of the learning process. The diverse quality and quantity of local training data mean that some nodes may initially exhibit poor classification performance. By sharing model updates, however, DFL has the potential to lift these underperforming nodes, leading to an overall improvement in network-wide anomaly detection.

In this paper, we leverage the Vehicular Reference Misbehavior (VeReMi) Extension Dataset \cite{VeremiExtension} to investigate the performance and security of DFL for safety message classification. We show that DFL can significantly improve classification accuracy compared to the local learning approach that relies solely on individual vehicle data. Vehicles that initially exhibit low individual accuracy due to limited local training samples experience gains when participating in DFL. We also uncover a strong correlation of overall model accuracy with local training data size and network connectivity over time.

Another critical consideration in adopting DFL is its vulnerability to multi-domain attacks. The decentralized nature of DFL, while reducing dependency on central infrastructure, also exposes the system to new security threats. In the wireless domain, jamming attacks may disrupt communication channels used for model propagation \cite{shi2022jamming,shi2023jamming, shi2022launch}. Concurrently, adversaries may launch data poisoning attacks, wherein compromised vehicles inject misleading information into the training process \cite{sagduyu2023securing, bataineh2024detecting, sagduyu2025}. DFL in a vehicular network setting is not immune to these attacks. Jamming attacks may prevent the exchange of model updates between vehicles, limiting the network coverage for DFL. Data poisoning attacks may contaminate local datasets to mislead the model, propagating misinformation and eventually degrading global performance. We show that jamming and poisoning attacks can be either individually or jointly launched to reduce the DFL accuracy effectively.  Our results uncover the characteristics of multi-domain threats to DFL and provide insights on vulnerabilities of DFL in practice. As vehicular networks become more interconnected, ensuring the robustness of DFL against multi-domain threats is of utmost importance. This paper makes the following contributions:
\begin{enumerate}
\item \textit{DFL Framework Design}: We apply DFL to anomaly detection in inter-vehicle networks, enabling one-hop model exchanges and multi-hop propagation without central server.
\item \textit{Performance Analysis}: Through extensive experiments with the VeReMi Extension Dataset, we demonstrate that DFL significantly outperforms local learning in terms of classification accuracy, benefiting nodes with limited data.
\item \textit{Vulnerability Characterization}: We perform a detailed analysis of DFL under multi-domain attacks, including  jamming and poisoning attacks, highlighting vulnerabilities of DFL to individual and joint attack effects.
\end{enumerate}

The remainder of the paper is organized as follows. Section~\ref{sec:sysmodel} introduces the system model for DFL in inter-vehicle networks. Section~\ref{sec:DFL} evaluates DFL performance for anomaly detection. Section~\ref{sec:attack} presents attacks on DFL. Section~\ref{sec:conclusion} concludes the paper.

\section{System Model}\label{sec:sysmodel}

\subsection{Vehicular Network Dataset}
The VeReMi Extension dataset was developed for malicious behavior detection in Cooperative Intelligent Transport Systems (C-ITS) \cite{VeremiExtension}. Each vehicle sends basic safety messages (BSMs) to the other vehicles in its communication range during its operation. This dataset is a collection of these BSMs and introduces both malfunctions (non-malicious errors due to faulty sensors) and attacks (intentional disruptions). Each message is tagged with type of message, timestamp of transmission and reception, sender identity, position, speed, acceleration, and heading (with and without noise). 

\begin{figure}[ht!]
    \centering
    \begin{subfigure}[b]{0.49\columnwidth}
        \centering
        \includegraphics[width=\linewidth]{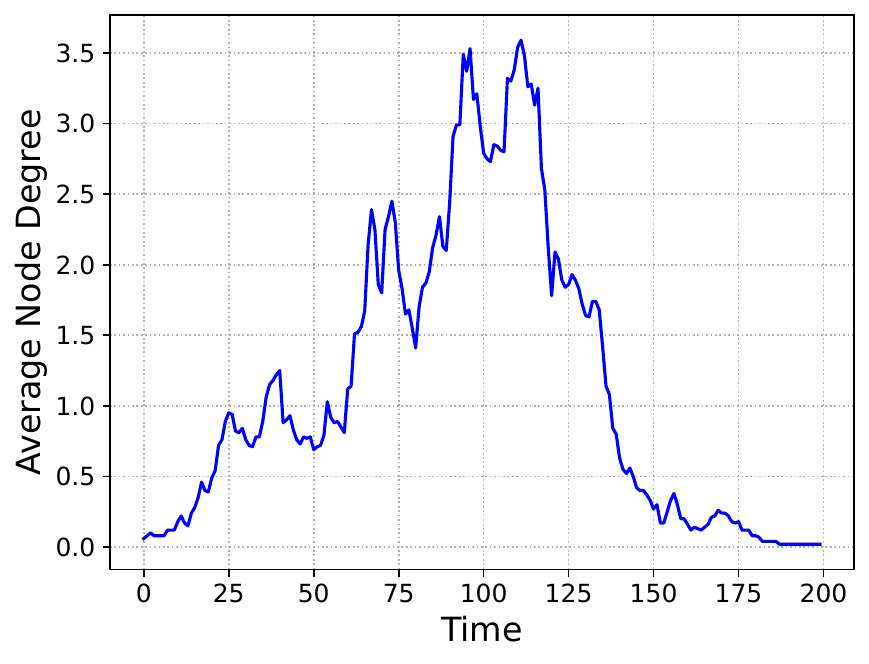}
        \caption{Average node degree.}
        \label{fig:node_degree}
    \end{subfigure}
    \hfill
    \begin{subfigure}[b]{0.49\columnwidth}
        \centering
        \includegraphics[width=\linewidth]{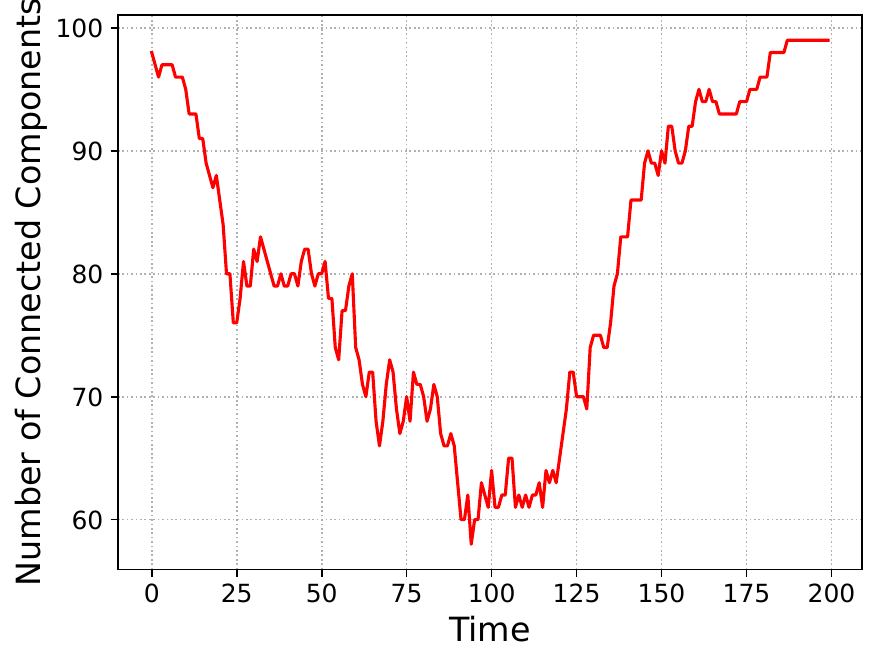}
        \caption{Connected components (CC).}
        \label{fig:num_cc}
    \end{subfigure}
    
    \begin{subfigure}[b]{0.49\columnwidth}
        \centering
        \includegraphics[width=\linewidth]{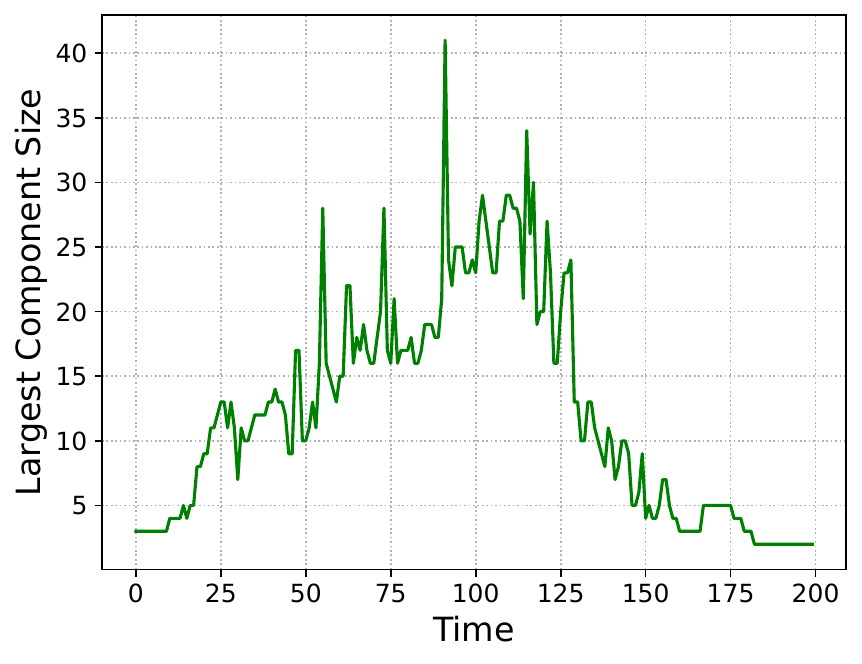}
        \caption{Largest CC size.}
        \label{fig:largest_cc}
    \end{subfigure}
    \hfill
    \begin{subfigure}[b]{0.49\columnwidth}
        \centering
        \includegraphics[width=\linewidth]{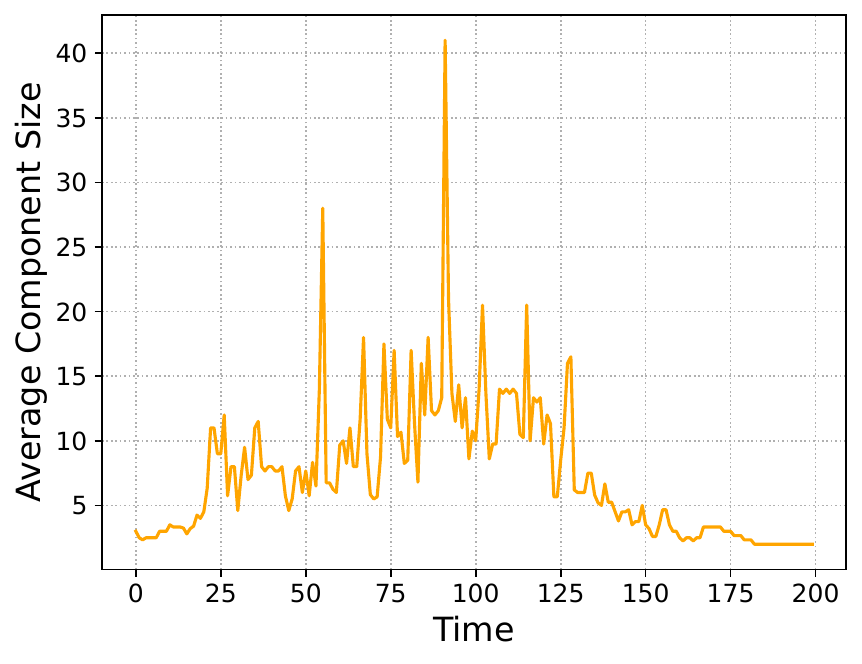}
        \caption{Average CC size.}
        \label{fig:average_cc}
    \end{subfigure}
    \caption{Network metrics over time.}
    \label{fig:all_metrics}
\end{figure}

\begin{figure}[t]
	\centering
	\includegraphics[width=0.67\columnwidth]{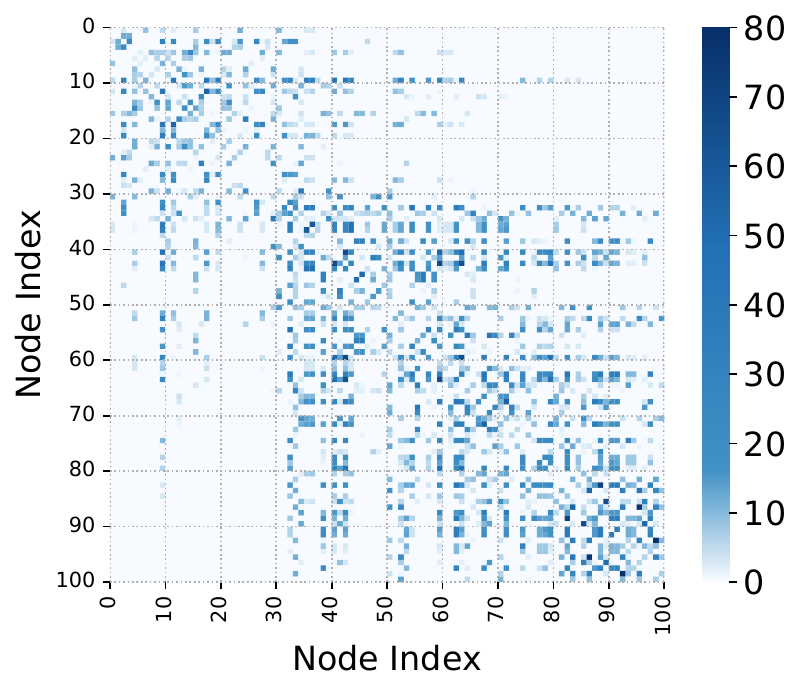}
	\caption{Aggregated adjacency matrix (link persistence).}
	\label{fig:connectivity}
\end{figure}

In this paper, we use the morning peak traffic dataset and focus on the Denial of Service (DoS) attacks in terms of high-frequency message flooding. There are a total of $64,779$ BSMs across all the users. We separate the data for training, validation and testing. We train a deep learning model at each vehicle using 22 features~\cite{ercanVeReMiDoS}. There are 100 vehicles (nodes) in the vehicular network, 94 of them with (training and test) data and 6 of them without data. The connectivity of nodes is determined by the adjacency matrix that changes over time with node mobility. Network metrics over time (average node degree, number of connected components, average connected component size, and largest connected component size) are shown in Figure~\ref{fig:all_metrics}. The heatmap for connectivity among nodes aggregated over time is shown in Figure~\ref{fig:connectivity}. We evaluate the relationship between the DFL performance and network properties in Section~\ref{sec:DFL}.

\subsection{Distributed Federated Learning Operation}
In FL, each vehicle trains a local model and shares only model updates instead of raw data. However, standard FL relies on a central server for model aggregation, which introduces bottlenecks and single points of failure. To address these challenges, DFL supports vehicles to exchange model updates directly with their one-hop neighbors in a peer-to-peer manner. Instead of requiring a central aggregator, models are aggregated locally and propagated through the network in a multi-hop fashion, enabling decentralized knowledge sharing while maintaining privacy and reducing reliance on fixed infrastructure. With DFL, each vehicle acts as an autonomous node that collects sensor data, processes it locally, and participates in collaborative learning. Main components of DFL training are:

\begin{enumerate}
\item \textit{Initial Local Training}: Each vehicle maintains a local dataset gathered from onboard sensors and communication modules (e.g., beacon messages, BSMs). Vehicles train deep learning models independently based on their local data.

\item \textit{Model Exchange with One-Hop Neighbors}:
Instead of sending data to a central server, each vehicle shares its locally trained models with one-hop neighbor vehicles.

\item \textit{Local Aggregation and Multi-Hop Propagation}: Each vehicle aggregates received model updates from its neighbors to its local model through federated averaging.
The updated model is propagated in a multi-hop manner across the network.

\item \textit{Iterative Learning}: Vehicles continue training and exchanging models over multiple rounds. Over time, this process allows all vehicles to collaboratively learn a global representation of the model without central coordination.

\end{enumerate}

\section{Anomaly Detection with Distributed Federated Learning} \label{sec:DFL}
In DFL, nodes collectively train their models to classify incoming messages as benign or malicious. 
As a benchmark, we consider local training only without model exchange.  For comparison purposes, we consider two types of DNN models, one small model and one large model, with architectures shown in Table~\ref{tab:small_large_dnn_architecture}. The nodes have training data sets ranging from 86 to 2514 samples, and the average size of the training data per node is 689 samples. On the other hand, nodes have test (inference) datasets ranging from 3 to 114 samples, and the average test data size per node is 36 samples.

\begin{figure}[t]
    \centering
    \begin{subfigure}[b]{0.49\columnwidth}
        \centering
        \includegraphics[width=\linewidth]{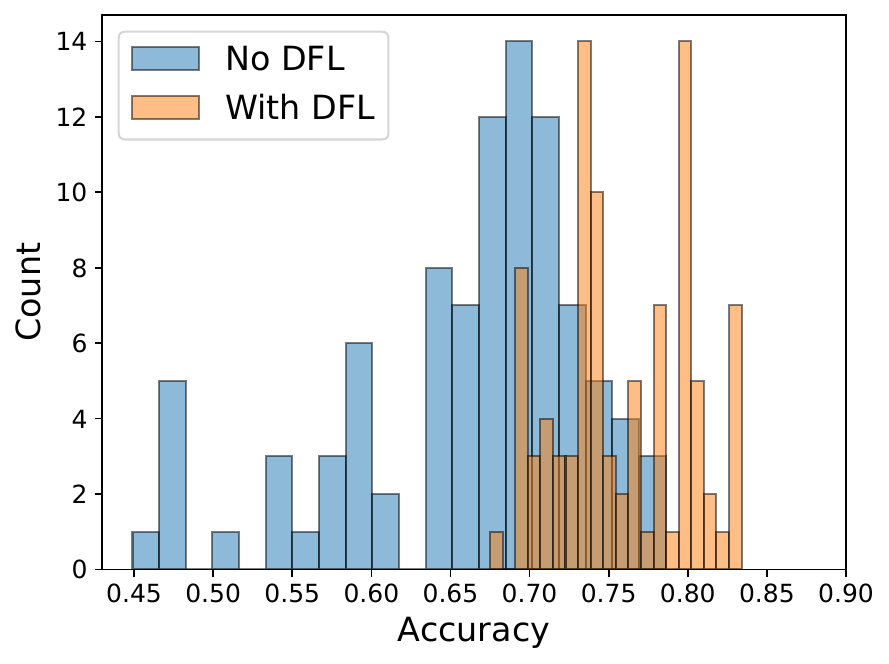} 
    \caption{When the small DNN is used.}
    \label{fig:Histogram_FL_noFL_Small} 
    \end{subfigure}
    \hfill
    \begin{subfigure}[b]{0.49\columnwidth}
        \centering
        \includegraphics[width=\linewidth]{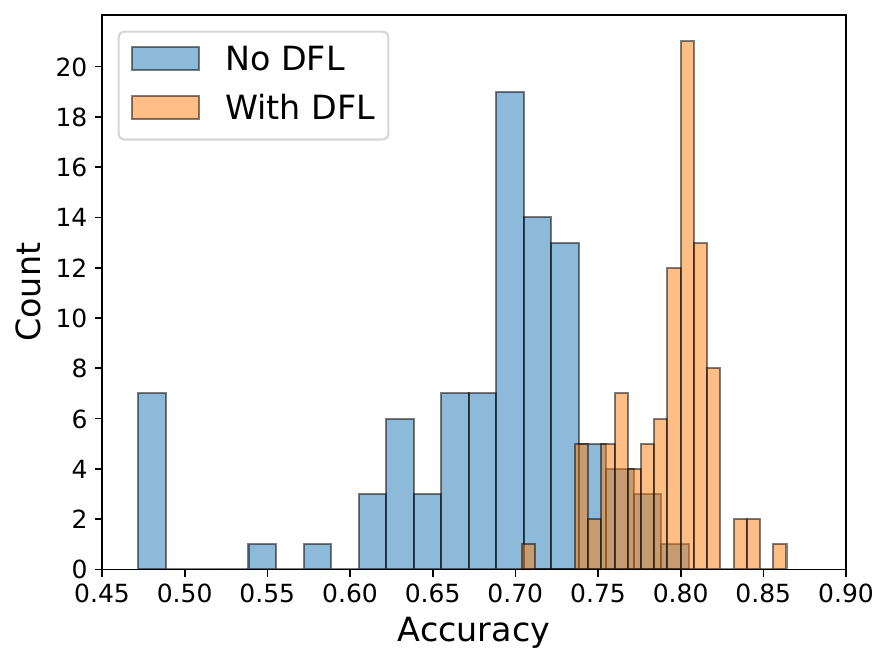} 
    \caption{When the large DNN is used.}
    \label{fig:Histogram_FL_noFL_Large} 
    \end{subfigure}
    \caption{Histogram of accuracies with and without DFL.}
    \label{fig:Histogram}
\end{figure}

We measure different accuracy metrics, including average, minimum, and maximum accuracy and its standard deviation across all nodes. Accuracy results are shown in Table~\ref{tab:small_large_nn}. The corresponding histograms of accuracy values achieved by all nodes are shown in Figure~\ref{fig:Histogram}. In both models, DFL improves every node’s classification accuracy, significantly raises the average and minimum accuracies, and reduces the standard deviation compared to local training alone. Overall, the large DNN model achieves higher accuracy compared to the small DNN model with DFL. Thus, for the rest of the paper, we continue with the large DNN model. 

\begin{table}[htbp]
\small
    \centering
    \caption{DNN architectures.}
    \label{tab:small_large_dnn_architecture}
        \centering
    \begin{tabular}{l|cc|cc}
        \toprule
        & \multicolumn{2}{c}{Small DNN} & \multicolumn{2}{c}{Large DNN}\\ 
        \hline
        Layer & Size & Activation & Size & Activation \\
        \hline
        Input     & 22  & --  & 22 & --\\
        Hidden 1  & 16  & ReLU & 128 & ReLU  \\
        Hidden 2  & 8   & ReLU  & 32 & ReLU\\
        Output    & 2   & Softmax & 2 & Softmax \\
        \bottomrule
    \end{tabular}
\end{table}

\begin{table}[htbp]
\small
    \centering
    \caption{Performance with and without DFL.}
    \label{tab:small_large_nn}
    \begin{subtable}[t]{0.45\textwidth}
        \centering
        \caption{Small DNN.}
        \label{tab:small_nn}
        \begin{tabular}{l|ccc}
        \toprule
        Metric & No DFL & With DFL & DFL Improvement \\
        \hline
        Average Accuracy & 0.6625 & 0.7592 & 14.60\% \\
        Minimum Accuracy & 0.4490 & 0.6750 & 50.33\% \\
        Maximum Accuracy & 0.7860 & 0.8339 & 6.09\% \\
        Standard Deviation & 0.0762 & 0.0426 & 44.09\% \\
        \bottomrule
        \end{tabular}
    \end{subtable}%
    
    \begin{subtable}[t]{0.45\textwidth}
        \centering
        \caption{Large DNN.}
        \label{tab:large_nn}
        \begin{tabular}{l|ccc}
        \toprule
        Metric & No DFL & With DFL & DFL Improvement \\
        \hline
        Average Accuracy & 0.6811 & 0.8004 & 17.52\% \\
        Minimum Accuracy & 0.4720 & 0.7040 & 49.15\% \\
        Maximum Accuracy & 0.7870 & 0.8640 & 9.78\% \\
        Standard Deviation & 0.0732 & 0.0275 & 62.43\% \\
        \bottomrule
        \end{tabular}
    \end{subtable}
\end{table}

Next, we analyze the correlation of the DFL accuracy with the local (individual) learning accuracy, the training data size, and the network properties including the node degree and the average size of connected components over time. We make the following definitions. $\mathcal{N}$ is the set of nodes with (training and test) data. $\boldsymbol{m}$ is the vector of training data sizes, where $\boldsymbol{m}_i$ is the training data size of node $i \in \mathcal{N}$. $\boldsymbol{a}_{\textit{DFL}}$ is the vector of DFL accuracies, where $\boldsymbol{a}_{\textit{DFL},i}$ is the DFL accuracy of node $i \in \mathcal{N}$. $\boldsymbol{a}_{\textit{LL}}$ is the vector of local learning accuracies without DFL, where $\boldsymbol{a}_{\textit{LL},i}$ is the local learning accuracy of node $i \in \mathcal{N}$.  $\boldsymbol{d}$ is the average incoming node degree, where $\boldsymbol{d}_i$ is the average number of incoming connections to that node $i \in \mathcal{N}$ over time. $\boldsymbol{C}$ is the average size of connected components for nodes, where $\boldsymbol{C}_i$ is the average size of connected components that node $i \in \mathcal{N}$ belongs to over time. $\boldsymbol{c}$ is the ratio of connected times for nodes, where $\boldsymbol{c}_i$ is the ratio of times when node $i \in \mathcal{N}$ is connected to at least one other node over time. 

\begin{table}[htbp]
\small
    \centering
    \caption{Correlation analysis.}
    \label{tab:pearson_correlation}
    \begin{tabular}{lc|lc}
        \toprule 
        \multicolumn{2}{c}{Pearson Correlation} & \multicolumn{2}{c}{Spearman's Rank Correlation}\\
        \hline 
        Terms & Coefficient & Terms & Coefficient \\
        \hline
        $\rho(\boldsymbol{a}_{\textit{DFL}}, \boldsymbol{a}_{\textit{LL}})$ & 0.3386 & $s(\boldsymbol{a}_{\textit{DFL}}, \boldsymbol{a}_{\textit{LL}})$ & 0.2895 \\
        $\rho(\boldsymbol{a}_{\textit{DFL}}, \boldsymbol{m})$ & 0.4365 & $s(\boldsymbol{a}_{\textit{DFL}}, \boldsymbol{m})$ & 0.4548 \\
        $\rho(\boldsymbol{a}_{\textit{LL}}, \boldsymbol{m})$ & 0.3074 &  $s(\boldsymbol{a}_{\textit{LL}}, \boldsymbol{m})$ & 0.3561  \\
        $\rho(\boldsymbol{a}_{\textit{DFL}}, \boldsymbol{d})$ &  0.4552 & $s(\boldsymbol{a}_{\textit{DFL}}, \boldsymbol{d})$ & 0.4988  \\
        $\rho(\boldsymbol{a}_{\textit{DFL}}, \boldsymbol{C})$ & 0.5771 & $s(\boldsymbol{a}_{\textit{DFL}}, \boldsymbol{C})$ &  0.6385 \\
        $\rho(\boldsymbol{a}_{\textit{DFL}}, \boldsymbol{c})$ & 0.4251 & $s(\boldsymbol{a}_{\textit{DFL}}, \boldsymbol{c})$ & 0.4808 \\
        \bottomrule
    \end{tabular}
\end{table}

Let $\rho(\boldsymbol{x}, \boldsymbol{y})$ and $s(\boldsymbol{x}, \boldsymbol{y})$ denote the Pearson correlation coefficient and Spearman's rank correlation coefficient, respectively, between $\boldsymbol{x}$ and $\boldsymbol{y}$. For the large DNN, the correlation analysis results are given in Table~\ref{tab:pearson_correlation} for Pearson correlation coefficient and Spearman's rank correlation. DFL accuracy and local learning accuracy are positively correlated with each other, and DFL accuracy has higher correlation with training data size compared to local learning accuracy. While local learning accuracy is not correlated with network properties in general, DFL accuracy is highly correlated with node degree, average size of connected components, and ratio of connected time. We utilize this correlation in Section~\ref{sec:attack} to define attacks on DFL.

\begin{figure}[t]
	\centering
	\includegraphics[width=0.9\columnwidth]{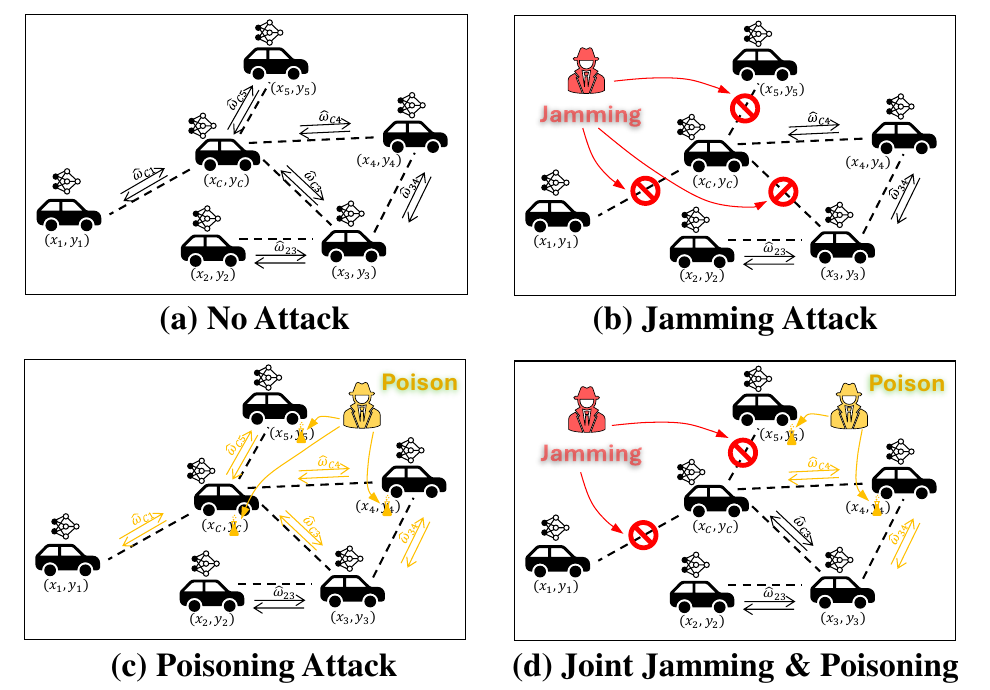}
	\caption{Jamming and poisoning attacks on DFL.}
	\label{fig:attack}
\end{figure}

\section{Multi-domain Attacks on Distributed Federated Learning} \label{sec:attack}

\subsection{Jamming Attacks}
\label{subsec:jamAttack}
Jamming attacks pose a significant threat to the performance and reliability of DFL in inter-vehicle networks. DFL relies on timely and repeated communication among neighbors to exchange and aggregate local model updates. Jamming can severely impair the propagation of learning updates (from Figure~\ref{fig:attack}a to Figure~\ref{fig:attack}b) and degrade the overall model quality. Nodes isolated by jamming cannot contribute their local knowledge to the network. Their neighbors miss out on potentially valuable updates, leading to incomplete or biased aggregation. Nodes affected by jamming can only continue local training, which degrades performance without DFL.

We assess DFL’s robustness to jamming attacks of varying intensity by selectively isolating a subset of nodes, i.e., severing their incoming connections, and measuring the resulting performance. In practice, jamming can be performed with directional transmissions without affecting the BSMs received at other vehicles. First, we select this group of nodes as the \textit{$Top K_{J}$} performers in the DFL learning scenario, where $J$ stands for jamming. The reason for this type of selection is that DFL allows users with multiple neighbors to exchange information (model weights) and learn more effectively. This also provides a higher attack success rate compared to random jamming. Later, we switch to selecting the set of jammed nodes based on the network connectivity properties. By gradually cutting the connections of the top performers in DFL, we eventually reduce the network down to local learning (no FL), which allows us to show how DFL enhances learning within the network.

We provide the results for average and minimum accuracies in Table~\ref{tab:jamming}. We start from DFL with no attacks and increase the number of users to jam from 10 to all users, which comes down to local learning, by effectively cutting all incoming connections to users. DFL achieves an average accuracy of $0.8004$ among all users within the network, where the minimum accuracy is $0.7034$. It is evident that as we cut the top performing users in DFL from the network and they can no longer provide their models to other participating nodes, the accuracy across the network decreases. When all the connections are cut, DFL is reduced to local learning with the average accuracy of $0.6811$ and minimum accuracy of 0.4720, underlining the impact of jamming on DFL performance.

\begin{table}[htbp]
\small
    \centering
    \caption{Effects of jamming attacks on DFL performance.}
    \label{tab:jamming}
    \begin{tabular}{ccc|ccc}
        \toprule
        \hspace{-0.1cm}Jammed   & Average  & Minimum & Jammed   & Average  & Minimum   \\
        \hspace{-0.1cm}Nodes & Accuracy & Accuracy & Nodes & Accuracy & Accuracy \\
        \hline
        \hspace{-0.1cm} None     & 0.8004 & 0.7034 & $Top 60_{J}$   & 0.7266 & 0.5199 \\
        \hspace{-0.1cm} $Top 10_{J}$   & 0.7976 & 0.6949& $Top 70_{J}$   & 0.7187 & 0.4749 \\
        \hspace{-0.1cm} $Top 20_{J}$   & 0.7763 & 0.6909& $Top 80_{J}$   & 0.7036 & 0.4729 \\
        \hspace{-0.1cm} $Top 30_{J}$   & 0.7690 & 0.6869& $Top 90_{J}$   & 0.6911 & 0.4729  \\
        \hspace{-0.1cm} $Top 40_{J}$   & 0.7680 & 0.6800& All      & 0.6811 & 0.4720 \\
        \hspace{-0.1cm} $Top 50_{J}$   & 0.7412 & 0.5600& -- & -- &--  \\
        \bottomrule
    \end{tabular}
\end{table}

\subsection{Poisoning Attacks}
\label{subsec:poisonAttack}
A training data poisoning attack is an adversarial machine learning attack that manipulates the training dataset by injecting misleading or malicious data samples to degrade the performance of a machine learning model. One common strategy is label manipulation, where the attacker intentionally assigns incorrect labels to training data samples  (Figure~\ref{fig:attack}c). Similar to the jamming case, we evaluate the performance of the DFL model using different poisoning intensities. 

We assume that the training data labels are flipped for a certain set of nodes with probability $p_a$. We select $Top K_{P}$ nodes to poison, where $P$ stands for poisoning. Some nodes cannot achieve high classification accuracy with only local learning. Thus, when their labels are flipped after poisoning, they may deceptively improve their classification accuracy. Some nodes already have high classification accuracy with local learning. Poisoning the training data samples of those nodes effectively reduces the overall classification accuracy. Contrary to the jamming case, we determine the set of users to poison from the best performers in the local learning (no DFL) case before the attack. While this process leads to an effective poisoning attack, it may not be possible for an attacker to know in advance which nodes to poison. Later, we replace this selection of poisoned nodes based on network connectivity properties that may be observed by the attacker over the air. For $Top K_{P}$, we select 47 as the half of the total number of nodes with data, and $25$ and $70$ as the intermediate node numbers. We select $p_a = \{0.25, 0.5, 0.75, 1\}$ as the probability of poisoning the labels of training data samples (from the $Top K_{P}$ nodes to be poisoned).  

We present poisoning attack results in Table~\ref{tab:poisoning_all} and compare them with the no-attack case. As we increase the poisoning attack intensity (higher $K$ in $Top K_{P}$, with higher probability of label flipping, $p_a$) average accuracies drop, eventually settling at basically random guessing, i.e., $0.5314$ when $p_a=1$ and top 70 high accuracy nodes, $Top70_{P}$, are poisoned. Similarly, minimum accuracy values drop significantly, even below a random guessing case, e.g., $0.3149$ when $p_a=1$ and top 70 best accuracy nodes, $Top70_{P}$, are poisoned. We observe that label poisoning attacks do not remain localized -- they propagate over multiple hops, gradually corrupting the learning process and leading to classification errors.
\begin{table}[htbp]
\small
    \centering
    \caption{Effects of poisoning attacks on DFL performance.}
    \label{tab:poisoning_all}
    \begin{subtable}[t]{0.45\textwidth}
        \centering
        \caption{Average accuracy.}
        \label{tab:poisoning_sub1}
        \begin{tabular}{c|c|c|c|c}
            \toprule
            Poisoned & \multicolumn{4}{c}{Poisoning Probability} \\
            Nodes    & $p_a = 0.25$ & $p_a = 0.5$ & $p_a = 0.75$ & $p_a = 1$ \\
            \hline
            None     &      \multicolumn{4}{c}{0.8004}    \\
            $Top 25_{P}$   &      0.7685  &     0.7532  &    0.7435   &   0.7378 \\
            $Top 47_{P}$   &      0.7577  &     0.7313  &    0.7309   &   0.6619 \\
            $Top 70_{P}$   &      0.7540  &     0.7166  &    0.6227   &   0.5314 \\
            \bottomrule
        \end{tabular}
    \end{subtable}
    \begin{subtable}[t]{0.45\textwidth}
        \centering
        \caption{Minimum accuracy.}
        \label{tab:poisoning_sub2}
        \begin{tabular}{c|c|c|c|c}
            \toprule
            Poisoned & \multicolumn{4}{c}{Poisoning Probability} \\
            Nodes    & $p_a = 0.25$ & $p_a = 0.5$ & $p_a = 0.75$ & $p_a = 1$ \\
            \hline
            None     &     \multicolumn{4}{c}{0.7040}     \\
            $Top 25_{P}$   &      0.6869    &     0.6779  &    0.6420   &   0.6480 \\
            $Top 47_{P}$   &      0.6869    &     0.6269  &    0.5320   &   0.4189 \\
            $Top 70_{P}$   &      0.6869    &     0.5939  &    0.4680   &   0.3149 \\
            \bottomrule
        \end{tabular}
    \end{subtable}

\end{table}

\subsection{Joint Jamming and Poisoning Attacks}
Next, we evaluate the effect of the simultaneous jamming and poisoning attacks (Figure~\ref{fig:attack}d) across a wide range of attack intensities. We use $p_a=\{0.5, 1\}$, $Top K_{J} = \{25, 47, 70\}$, and $Top K_{P} = \{25, 47, 70\}$ in the simulations. Results are provided in Table~\ref{tab:jam_and_poison}.  The average accuracy drops within each subtable as the jamming intensity increases, i.e., higher $Top K_{J}$, and all of these accuracies are lower than the no attack DFL accuracy of $80\%$. Within each subtable higher $p_a$ results in lower accuracy, as expected.

By comparing Table~\ref{tab:jam_poison_sub1a} and Table~\ref{tab:jam_poison_sub2a}, we see that when more $Top K_{P}$ nodes are poisoned, overall accuracy consistently decreases and the accuracies in Table~\ref{tab:jam_poison_sub2a} are consistently lower than their counterparts in Table~\ref{tab:jam_poison_sub1a}. However, when we compare Table~\ref{tab:jam_poison_sub2a} with Table~\ref{tab:jam_poison_sub3a}, as $Top K_{J}$ nodes are jammed, the resulting accuracies are higher compared to their counterparts in Table~\ref{tab:jam_poison_sub2a}. The reason is that as more nodes are included in $Top K_{P}$, it is more likely that the nodes with low accuracy will be poisoned, as well. Since labels are binary, when labels of those nodes are flipped, their classification accuracy may improve. When nodes with low classification accuracy are jammed, i.e., cut from the network, their potentially misleading model weights can no longer affect the network during DFL. This results in higher classification accuracy.

\subsection{Joint Attacks based on Network Properties}
Lastly, we build upon the correlation of node accuracies and network properties that we have established in the correlation analysis of Section~\ref{sec:DFL}, and consider attacks on the DFL with additional sets of attacked nodes, i.e., $Top K_{d}$, $Top K_{C}$, and $Top K_{c}$, where we order nodes with respect to node degree, average connected component size, and connection time ratio. We test the performance with $K = \{25, 47, 70\}$ for individual jamming and poisoning, as well as combined attack, where we choose $p_a = \{0.5, 1\}$. We provide the results in Table~\ref{tab:March2526Table}. In all three subtables,  as the attack intensity increases, accuracy drops within each attack type, as expected. We also observe that poisoning attack is more impactful compared to jamming, where the joint attack lowers the accuracies even further. As also observed in Table~\ref{tab:jam_and_poison}, the effects of high degrees of joint jamming and poisoning attacks may cancel each other (with $p_a=1$ across all $K$ values), resulting in slightly higher accuracies compared to poisoning only.

\begin{table}[htbp]
\small
    \centering
    \caption{Joint effects of poisoning and jamming attacks.}
    \label{tab:jam_and_poison}
    \begin{subtable}[t]{0.21\textwidth}
        \centering
        \caption{Average accuracy.}
        \label{tab:jam_poison_sub1a}
        \begin{tabular}{c|c|c}
            \toprule
            Jammed & \multicolumn{2}{c}{$Top 25_{P}$ Poison Prob.} \\
            Nodes    & $p_a = 0.5$ & $p_a = 1$ \\
            \hline
            None     &     0.7532 & 0.7378  \\
            $Top 25_{J}$   &      0.7342  &     0.7234  \\
            $Top 47_{J}$   &      0.7168  &     0.7040  \\
            $Top 70_{J}$   &      0.6951  &     0.6652  \\
            \bottomrule
        \end{tabular}
    \end{subtable}%
    \hspace{0.75cm}
    \begin{subtable}[t]{0.21\textwidth}
        \centering
        \caption{Minimum accuracy.}
        \label{tab:jam_poison_sub1b}
        \begin{tabular}{c|c|c}
            \toprule
            Jammed & \multicolumn{2}{c}{$Top 25_{P}$ Poison Prob.} \\
            Nodes    & $p_a = 0.5$ & $p_a = 1$ \\
            \hline
            None     &      0.6779  &  0.6480    \\
            $Top 25_{J}$   &      0.4729  &     0.4740  \\
            $Top 47_{J}$   &      0.5559  &     0.3740  \\
            $Top 70_{J}$   &      0.4760  &     0.3310  \\
            \bottomrule
        \end{tabular}
    \end{subtable}
        \begin{subtable}[t]{0.21\textwidth}
        \centering
        \caption{Average accuracy.}
        \label{tab:jam_poison_sub2a}
        \begin{tabular}{c|c|c}
            \toprule
            Jammed & \multicolumn{2}{c}{$Top 47_{P}$ Poison Prob.} \\
            Nodes    & $p_a = 0.5$ & $p_a = 1$ \\
            \hline
            None     &     0.7313 & 0.6619    \\
            $Top 25_{J}$   &      0.6002  &     0.4837  \\
            $Top 47_{J}$   &      0.5928  &     0.4719  \\
            $Top 70_{J}$   &      0.6048  &     0.4726  \\
            \bottomrule
        \end{tabular}
    \end{subtable}
    \hspace{0.75cm}
    \begin{subtable}[t]{0.21\textwidth}
        \centering
        \caption{Minimum accuracy.}
        \label{tab:jam_poison_sub2b}
        \begin{tabular}{c|c|c}
            \toprule
            Jammed & \multicolumn{2}{c}{$Top 47_{P}$ Poison Prob.} \\
            Nodes    & $p_a = 0.5$ & $p_a = 1$ \\
            \hline
            None     &      0.6269 & 0.4189  \\
            $Top 25_{J}$   &      0.3460  &     0.2700  \\
            $Top 47_{J}$   &      0.3550  &     0.2390  \\
            $Top 70_{J}$   &      0.3920  &     0.2640  \\
            \bottomrule
        \end{tabular}
    \end{subtable}
        \begin{subtable}[t]{0.21\textwidth}
        \centering
        \caption{Average accuracy.}
        \label{tab:jam_poison_sub3a}
        \begin{tabular}{c|c|c}
            \toprule
            Jammed & \multicolumn{2}{c}{$Top 70_{P}$ Poison Prob.} \\
            Nodes    & $p_a = 0.5$ & $p_a = 1$ \\
            \hline
            None     &     0.7166 & 0.5314    \\
            $Top 25_{J}$   &      0.7081  &     0.5276  \\
            $Top 47_{J}$   &      0.6837  &     0.5622  \\
            $Top 70_{J}$   &      0.6657  &     0.5264  \\
            \bottomrule
        \end{tabular}
    \end{subtable}
    \hspace{0.75cm}
    \begin{subtable}[t]{0.21\textwidth}
        \centering
        \caption{Minimum accuracy.}
        \label{tab:jam_poison_sub3b}
        \begin{tabular}{c|c|c}
            \toprule
            Jammed & \multicolumn{2}{c}{$Top 70_{P}$ Poison Prob.} \\
            Nodes    & $p_a = 0.5$ & $p_a = 1$ \\
            \hline
            None     &      0.5939  & 0.3149   \\
            $Top 25_{J}$   &      0.5469  &     0.3490  \\
            $Top 47_{J}$   &      0.4679  &     0.3249  \\
            $Top 70_{J}$   &      0.4729  &     0.3170  \\
            \bottomrule
        \end{tabular}
    \end{subtable}
\end{table}

\begin{table}[htbp]
\small
    \centering
    \caption{Effects of attacks based on network properties.}
    \label{tab:March2526Table}
    \begin{subtable}[t]{0.5\textwidth}
        \centering
        \caption{Node degree.}
        \label{tab:March2526Table_sub1}
    \begin{tabular}{l|cc|cc|cc}
        \toprule
        & \multicolumn{2}{c}{Jam Only} & \multicolumn{2}{c}{Poison Only} & \multicolumn{2}{c}{Joint Attack}\\ 
        \hline
        Attack param. & Avg. & Min & Avg. & Min & Avg. & Min \\
        \hline
        $Top 25_{d}, p_a=0.5$     
         & \multirow{2}{*}{0.7651}  
         & \multirow{2}{*}{0.6750}  
         & 0.7722 & 0.6769 & 0.7598 & 0.7080 \\
        $Top 25_{d}, p_a=1.0$     
         &                         &                         & 0.7204 & 0.4779 & 0.7260 & 0.5310 \\
        $Top 47_{d}, p_a=0.5$     
         & \multirow{2}{*}{0.7353}  
         & \multirow{2}{*}{0.4740}  
         & 0.7307 & 0.6430 & 0.7261 & 0.6380 \\
        $Top 47_{d}, p_a=1.0$     
         &                         &                         & 0.6407 & 0.3449 & 0.6494 & 0.3790 \\
        $Top 70_{d}, p_a=0.5$     
         & \multirow{2}{*}{0.6903}  
         & \multirow{2}{*}{0.4480}  
         & 0.7070 & 0.5950 & 0.6903 & 0.4480 \\
        $Top 70_{d}, p_a=1.0$     
         &                         &                         & 0.5360 & 0.3190 & 0.5599 & 0.2720 \\
        \bottomrule
    \end{tabular}
    \end{subtable}
    \\
    \hspace{0.75cm}
    \begin{subtable}[t]{0.5\textwidth}
        \centering
        \caption{Average connected component size.}
        \label{tab:March2526Table_sub2}
    \begin{tabular}{l|cc|cc|cc}
        \toprule
        & \multicolumn{2}{c}{Jam Only} & \multicolumn{2}{c}{Poison Only} & \multicolumn{2}{c}{Joint Attack}\\ 
        \hline
        Attack param. & Avg. & Min & Avg. & Min & Avg. & Min \\
        \hline
        $Top 25_{C}, p_a=0.5$     
         & \multirow{2}{*}{0.7613}  
         & \multirow{2}{*}{0.6570}  
         & 0.7560 & 0.6710 & 0.7527 & 0.6589 \\
        $Top 25_{C}, p_a=1.0$     
         &   &   & 0.7008 & 0.3899 & 0.7188 & 0.4359 \\
        $Top 47_{C}, p_a=0.5$     
         & \multirow{2}{*}{0.7435}  
         & \multirow{2}{*}{0.6520}  
         & 0.7246 & 0.4729 & 0.7114 & 0.4760 \\
        $Top 47_{C}, p_a=1.0$     
         &   &   & 0.6102 & 0.3230 & 0.6339 & 0.3700 \\
        $Top 70_{C}, p_a=0.5$     
         & \multirow{2}{*}{0.7088}  
         & \multirow{2}{*}{0.4740}  
         & 0.7148 & 0.6050 & 0.6914 & 0.4650 \\
        $Top 70_{C}, p_a=1.0$     
         &   &   & 0.5466 & 0.2969 & 0.5434 & 0.3249 \\
        \bottomrule
    \end{tabular}
    \end{subtable}
        \\
    \hspace{0.75cm}
    \begin{subtable}[t]{0.5\textwidth}
        \centering
        \caption{Connected time ratio.}
        \label{tab:March2526Table_sub3}
    \begin{tabular}{l|cc|cc|cc}
        \toprule
        & \multicolumn{2}{c}{Jam Only} & \multicolumn{2}{c}{Poison Only} & \multicolumn{2}{c}{Joint Attack}\\ 
        \hline
        Attack param. & Avg. & Min & Avg. & Min & Avg. & Min \\
        \hline
        $Top 25_{c}, p_a=0.5$     
         & \multirow{2}{*}{0.7583}  
         & \multirow{2}{*}{0.4749}  
         & 0.7471 & 0.6819 & 0.7637 & 0.6880 \\
        $Top 25_{c}, p_a=1.0$     
         &   &   & 0.7316 & 0.6729 & 0.7505 & 0.6869 \\
        $Top 47_{c}, p_a=0.5$     
         & \multirow{2}{*}{0.7390}  
         & \multirow{2}{*}{0.5979}  
         & 0.7340 & 0.6460 & 0.7131 & 0.4140 \\
        $Top 47_{c}, p_a=1.0$     
         &   &   & 0.6429 & 0.4410 & 0.6658 & 0.3799 \\
        $Top 70_{c}, p_a=0.5$     
         & \multirow{2}{*}{0.7168}  
         & \multirow{2}{*}{0.4760}  
         & 0.7066 & 0.4729 & 0.6971 & 0.4530 \\
        $Top 70_{c}, p_a=1.0$     
         &   &   & 0.5395 & 0.3109 & 0.5555 & 0.2879 \\
        \bottomrule
    \end{tabular}
    \end{subtable}

\end{table}

\section{Conclusion} \label{sec:conclusion}
We explored the distributed operation of FL for anomaly detection in inter-vehicle networks, highlighting its advantages, challenges, and vulnerabilities in the presence of multi-domain attacks such as training data poisoning and  jamming. Unlike traditional FL, which relies on a central server for aggregation, DFL enables vehicles to train and exchange models locally with their one-hop neighbors, propagating knowledge through multi-hop communications.  Using the VeReMi Extension Dataset, we showed that DFL improves anomaly detection compared to local-only learning, while network connectivity and data heterogeneity strongly influence model accuracy. A major concern in DFL is its vulnerability to multi-domain attacks. We showed how training data poisoning attacks propagate over multiple hops, affecting not only the initially compromised vehicles but also indirectly other ones through repeated model aggregation, and causing network-wide degradation. On the other hand, jamming attacks can disrupt model exchanges, leading to incomplete learning. We evaluated individual and joint effects of jamming and poisoning attacks. Our results highlight that ensuring adversarial robustness and resilient communication is essential for safety and reliability of AI-driven inter-vehicle networks. Future work can investigate other attacks, such as targeted poisoning and backdoor attacks, and design defense mechanisms.

\section*{Acknowledgments}
Research was sponsored by the Army Research Laboratory under RTX BBN Technologies, Inc. subcontract and was accomplished under Cooperative Agreement Number W911NF-24-2-0131. The views and conclusions contained in this document are those of the authors and should not be interpreted as representing the official policies, either expressed or implied, of the Army Research Laboratory or the U.S. Government. The U.S. Government is authorized to reproduce and distribute reprints for Government purposes notwithstanding any copyright notation herein.

We would like to thank Stephen Raio from U.S. Army DEVCOM Army Research Laboratory for valuable feedback and guidance. 
\bibliographystyle{ACM-Reference-Format}
\bibliography{references}


\begin{thebibliography}{14}


\ifx \showCODEN    \undefined \def \showCODEN     #1{\unskip}     \fi
\ifx \showDOI      \undefined \def \showDOI       #1{#1}\fi
\ifx \showISBNx    \undefined \def \showISBNx     #1{\unskip}     \fi
\ifx \showISBNxiii \undefined \def \showISBNxiii  #1{\unskip}     \fi
\ifx \showISSN     \undefined \def \showISSN      #1{\unskip}     \fi
\ifx \showLCCN     \undefined \def \showLCCN      #1{\unskip}     \fi
\ifx \shownote     \undefined \def \shownote      #1{#1}          \fi
\ifx \showarticletitle \undefined \def \showarticletitle #1{#1}   \fi
\ifx \showURL      \undefined \def \showURL       {\relax}        \fi
\providecommand\bibfield[2]{#2}
\providecommand\bibinfo[2]{#2}
\providecommand\natexlab[1]{#1}
\providecommand\showeprint[2][]{arXiv:#2}

\bibitem[Bataineh et~al\mbox{.}(2024)]%
        {bataineh2024detecting}
\bibfield{author}{\bibinfo{person}{Ahmed~Saleh Bataineh}, \bibinfo{person}{Mohammad Zulkernine}, \bibinfo{person}{Adel Abusitta}, {and} \bibinfo{person}{Talal Halabi}.} \bibinfo{year}{2024}\natexlab{}.
\newblock \showarticletitle{Detecting Poisoning Attacks in Collaborative {IDS}s of Vehicular Networks Using XAI and Shapley Value}.
\newblock \bibinfo{journal}{\emph{Journal on Autonomous Transportation Systems}} \bibinfo{volume}{2}, \bibinfo{number}{3} (\bibinfo{year}{2024}).
\newblock


\bibitem[Ercan et~al\mbox{.}(2023)]%
        {ercanVeReMiDoS}
\bibfield{author}{\bibinfo{person}{Secil Ercan}, \bibinfo{person}{Leo Mendiboure}, \bibinfo{person}{Lylia Alouache}, \bibinfo{person}{Sassi Maaloul}, \bibinfo{person}{Tidiane Sylla}, {and} \bibinfo{person}{Hasnaa Aniss}.} \bibinfo{year}{2023}\natexlab{}.
\newblock \showarticletitle{An Enhanced Model for Machine Learning-Based DoS Detection in Vehicular Networks}. In \bibinfo{booktitle}{\emph{IFIP Networking Conference (IFIP Networking)}}.
\newblock


\bibitem[Huang et~al\mbox{.}(2024)]%
        {huang2024semi}
\bibfield{author}{\bibinfo{person}{Jiaqi Huang}, \bibinfo{person}{Yili Jiang}, \bibinfo{person}{Sohan Gyawali}, \bibinfo{person}{Zhiguo Zhou}, {and} \bibinfo{person}{Fangtian Zhong}.} \bibinfo{year}{2024}\natexlab{}.
\newblock \showarticletitle{Semi-supervised Federated Learning for Misbehavior Detection of {BSMs} in Vehicular Networks}. In \bibinfo{booktitle}{\emph{IEEE 100th Vehicular Technology Conference (VTC2024-Fall)}}.
\newblock


\bibitem[Kamel et~al\mbox{.}(2020)]%
        {VeremiExtension}
\bibfield{author}{\bibinfo{person}{Joseph Kamel}, \bibinfo{person}{Michael Wolf}, \bibinfo{person}{Rens~W. van~der Hei}, \bibinfo{person}{Arnaud Kaiser}, \bibinfo{person}{Pascal Urien}, {and} \bibinfo{person}{Frank Kargl}.} \bibinfo{year}{2020}\natexlab{}.
\newblock \showarticletitle{VeReMi Extension: A Dataset for Comparable Evaluation of Misbehavior Detection in {VANETs}}. In \bibinfo{booktitle}{\emph{IEEE International Conference on Communications (ICC)}}.
\newblock


\bibitem[Lu et~al\mbox{.}(2022)]%
        {9645261}
\bibfield{author}{\bibinfo{person}{Qiang Lu}, \bibinfo{person}{Hojin Jung}, {and} \bibinfo{person}{Kyoung-Dae Kim}.} \bibinfo{year}{2022}\natexlab{}.
\newblock \showarticletitle{Optimization-Based Approach for Resilient Connected and Autonomous Intersection Crossing Traffic Control Under V2X Communication}.
\newblock \bibinfo{journal}{\emph{IEEE Transactions on Intelligent Vehicles}} \bibinfo{volume}{7}, \bibinfo{number}{2} (\bibinfo{year}{2022}).
\newblock


\bibitem[Sagduyu et~al\mbox{.}(2023)]%
        {sagduyu2023securing}
\bibfield{author}{\bibinfo{person}{Yalin~E Sagduyu}, \bibinfo{person}{Tugba Erpek}, {and} \bibinfo{person}{Yi Shi}.} \bibinfo{year}{2023}\natexlab{}.
\newblock \showarticletitle{Securing {NextG} Systems against Poisoning Attacks on Federated Learning: A Game-Theoretic Solution}. In \bibinfo{booktitle}{\emph{IEEE Military Communications Conference (MILCOM)}}.
\newblock


\bibitem[Sagduyu et~al\mbox{.}(2025)]%
        {sagduyu2025}
\bibfield{author}{\bibinfo{person}{Y.~E. Sagduyu}, \bibinfo{person}{T. Erpek}, {and} \bibinfo{person}{Y. Shi}.} \bibinfo{year}{2025}\natexlab{}.
\newblock \showarticletitle{Poisoning Attack and Defense Game for Federated Learning in Resilient {NextG} Networks}.
\newblock In \bibinfo{booktitle}{\emph{Autonomous Cyber Resilience}}. \bibinfo{publisher}{Wiley-IEEE Press}.
\newblock
\newblock
\shownote{to appear}.


\bibitem[Shi and Sagduyu(2022a)]%
        {shi2022launch}
\bibfield{author}{\bibinfo{person}{Yi Shi} {and} \bibinfo{person}{Yalin~E Sagduyu}.} \bibinfo{year}{2022}\natexlab{a}.
\newblock \showarticletitle{How to launch jamming attacks on federated learning in NextG wireless networks}. In \bibinfo{booktitle}{\emph{IEEE Globecom Workshops (GC Wkshps)}}.
\newblock


\bibitem[Shi and Sagduyu(2022b)]%
        {shi2022jamming}
\bibfield{author}{\bibinfo{person}{Yi Shi} {and} \bibinfo{person}{Yalin~E Sagduyu}.} \bibinfo{year}{2022}\natexlab{b}.
\newblock \showarticletitle{Jamming attacks on federated learning in wireless networks}.
\newblock \bibinfo{journal}{\emph{arXiv preprint arXiv:2201.05172}} (\bibinfo{year}{2022}).
\newblock


\bibitem[Shi et~al\mbox{.}(2022)]%
        {shi2022federated}
\bibfield{author}{\bibinfo{person}{Yi Shi}, \bibinfo{person}{Yalin~E Sagduyu}, {and} \bibinfo{person}{Tugba Erpek}.} \bibinfo{year}{2022}\natexlab{}.
\newblock \showarticletitle{Federated learning for distributed spectrum sensing in {NextG} communication networks}.
\newblock \bibinfo{journal}{\emph{arXiv preprint arXiv:2204.03027}} (\bibinfo{year}{2022}).
\newblock


\bibitem[Shi et~al\mbox{.}(2023)]%
        {shi2023jamming}
\bibfield{author}{\bibinfo{person}{Yi Shi}, \bibinfo{person}{Yalin~E Sagduyu}, {and} \bibinfo{person}{Tugba Erpek}.} \bibinfo{year}{2023}\natexlab{}.
\newblock \showarticletitle{Jamming attacks on decentralized federated learning in general multi-hop wireless networks}. In \bibinfo{booktitle}{\emph{IEEE Conference on Computer Communications Workshops (INFOCOM WKSHPS)}}.
\newblock


\bibitem[Tham et~al\mbox{.}(2023)]%
        {tham2023federated}
\bibfield{author}{\bibinfo{person}{Chen-Khong Tham}, \bibinfo{person}{Lu Yang}, \bibinfo{person}{Akshit Khanna}, {and} \bibinfo{person}{Bhavya Gera}.} \bibinfo{year}{2023}\natexlab{}.
\newblock \showarticletitle{Federated learning for anomaly detection in vehicular networks}. In \bibinfo{booktitle}{\emph{IEEE 97th Vehicular Technology Conference (VTC2023-Spring)}}.
\newblock


\bibitem[Wang and Yan(2023)]%
        {wang2023federated}
\bibfield{author}{\bibinfo{person}{Zhe Wang} {and} \bibinfo{person}{Tingkai Yan}.} \bibinfo{year}{2023}\natexlab{}.
\newblock \showarticletitle{Federated learning-based vehicle trajectory prediction against cyberattacks}. In \bibinfo{booktitle}{\emph{IEEE 29th International Symposium on Local and Metropolitan Area Networks (LANMAN)}}.
\newblock


\bibitem[Yakan et~al\mbox{.}(2023)]%
        {yakan2023federated}
\bibfield{author}{\bibinfo{person}{Hadi Yakan}, \bibinfo{person}{Ilhem Fajjari}, \bibinfo{person}{Nadjib Aitsaadi}, {and} \bibinfo{person}{Cedric Adjih}.} \bibinfo{year}{2023}\natexlab{}.
\newblock \showarticletitle{Federated learning for {V2X} misbehavior detection system in {5G} edge networks}. In \bibinfo{booktitle}{\emph{ACM Conference on Modeling Analysis and Simulation of Wireless and Mobile Systems}}.
\newblock


\end{thebibliography}
	
\end{document}